\documentclass[12pt,a4paper,epsf]{article}
\usepackage{graphics}
\usepackage{amssymb,amsmath}
\usepackage[dvips]{lscape,graphicx}
\usepackage{cite}
\usepackage{longtable}

\newcommand{\ct}{\cite}
\newcommand{\lb}{\label}

\newcommand{\bc}{\begin{center}}
\newcommand{\ec}{\end{center}}
\newcommand{\bd}{\begin{displaymath}}
\newcommand{\ed}{\end{displaymath}}
\newcommand{\be}{\begin{equation}}
\newcommand{\ee}{\end{equation}}
\newcommand{\ba}{\begin{array}}
\newcommand{\ea}{\end{array}}
\newcommand{\bea}{\begin{eqnarray}}
\newcommand{\eea}{\end{eqnarray}}
\newcommand{\bt}{\begin{tabular}}
\newcommand{\et}{\end{tabular}}

\newcommand{\ov}{\overline}
\newcommand{\bp}{\begin{picture}}
\newcommand{\ep}{\end{picture}}
\newcommand{\bfi}{\begin{figure}}
\newcommand{\efi}{\end{figure}}

\def\fun#1#2{\lower3.6pt\vbox{\baselineskip0pt\lineskip.9pt
\ialign{$\mathsurround=0pt#1\hfil##\hfil$\crcr#2\crcr\sim\crcr}}}

\begin{document}

\title{\Large\bf Seesaw scales and Steps from the Standard Model
towards Superstring-Inspired Flipped $E_6$}
\author{\large C.R.~Das
 \footnote{\large\, crdas@imsc.res.in}\,\,
and Larisa Laperashvili ${}^{1,\, 2}$ \footnote{\large\, laper@itep.ru,
laper@imsc.res.in}\,\,,\\[5mm]
\itshape{${}^{1}$  The Institute of Mathematical Sciences,
Chennai, India}
\\[5mm]
\itshape{${}^{2}$ The Institute of Theoretical and Experimental Physics, Moscow,
 Russia}}

\date{}
\maketitle


\thispagestyle{empty}

\begin{abstract}

Recently in connection with Superstring theory $E_8$ and $E_6$
unifications became very promising. In the present paper we have
investigated a number of available paths from the Standard Model
(SM) to the $E_6$ unification, considering a chain of flipped
models following the extension of the SM:
$$
 SU(3)_C\times SU(2)_L\times U(1)_Y \to
 SU(3)_C\times SU(2)_L\times U(1)_X \times U(1)_Z
\to
$$ $$SU(5)\times U(1)_X \to  SU(5)\times U(1)_{Z1}
\times U(1)_{X1} \to SO(10) \times U(1)_{X1} \to $$ $$ SO(10)
\times U(1)_{Z2}\times U(1)_{X2} \to  E_6\times U(1)_{X2}\quad
{\mbox{or}}\quad E_6,
$$
Also we have considered a chain with a left-right symmetry:
$$
SU(3)_C\times SU(2)_L\times U(1)_Y \to SU(3)_C\times SU(2)_L
\times SU(2)_R\times U(1)_X\times U(1)_Z$$ $$ \to SU(4)_C\times
SU(2)_L \times SU(2)_R\times U(1)_Z$$ $$ \to SO(10)\times U(1)_Z
\to E_6.
$$
We have presented four examples including non-supersymmetric and
supersymmetric extensions of the SM and different contents of the
Higgs bosons providing the breaking of the flipped $SO(10)$ and
$SU(5)$ down to the SM. It was shown that the final unification
$E_6\times U(1)$ or $E_6$ at the (Planck) GUT scale $M_{SSG}$
depends on the number of the Higgs boson representations
considered in theory.
\end{abstract}

\clearpage\newpage

\pagenumbering{arabic}

\vspace{1cm}

\section{Introduction}

Nature is described by gauge theories and prefers to have gauge
groups with small rank, in comparison with the number $N_f$ of
given fermion fields. For instance, the Standard Model (SM)
contains 45 known fermion fields and could be described by the
enormous global group $SU(45)\times U(1)$. Why does Nature choose
only small subgroups of this enormous global group? The answer was
given in Refs.~\ct{1,2,3,4}. The principles are simple: the
resulting theory has to be (i) free from anomalies and (ii) free
from bare masses.  The largest simple subgroups of $SU(N_f)\times
U(1)$ satisfying conditions (i) and (ii) are listed for a given
$N_f$ in the Table 1 of Ref.~\ct{2} with $N_f$ going up to
$N_f=81$. If there exists one right-handed neutrino per family, in
addition to the known SM fermions, then with one, two or three
families we have the number of fermions 16, 32, 48, respectively.
For example, we have $N_f=16$ for $SU(5)$, $SO(10)$, $E_6$, and
$N_f=48$ for $[SU(5)]^3=SU(5)\times SU(5) \times SU(5)$,
$[SO(10)]^3$, $[E_6]^3$, etc., so called family replicated gauge
group models \ct{4a}.

The last 10 years Grand Unified Theories (GUTs) are inspired by
superstring theories \ct{5,6,7,8,9} showing a connection between
superstrings and low energy particle physics phenomenology. The
most realistic model based on the string theory is the heterotic
string-derived flipped $E_6$ \ct{10,11}. Compactification theory
involving D-branes provides interesting features of the flipped
models. Conventional GUT models, such as $SU(5)$ and $SO(10)$,
have been investigated in details by a lot of authors, but none of
them are not completely satisfactory. In Refs.~\ct{10} the gauge
group $SU(5)\times U(1)$ ('flipped $SU(5)$') was suggested as a
very economical candidate for unified theory: it may require only
$10+\ov {10}$ Higgs representations to break the GUT symmetry, in
contrast to other unified models which require large
representations. There are many attractive experimentally testable
results for flipped $SU(5)$, including the prediction of
$\alpha_s(M_Z)$ \ct{12}, doublet-triplet mass splitting mechanism
\ct{10,12a}, seesaw neutrino mass relations \ct{12a}, explanation
of the leptophobic $Z'$ phenomenon \ct{13}. The hierarchy problem
between the electroweak Higgs doublets and the colour Higgs
triplets is solved giving a natural suppression of 5-dimensional
operators that may mediate rapid proton decay. For this reason
flipped $SU(5)$ is in agreement with experimental limits placed
upon proton lifetime \ct{14} (see also \ct{14a} and \ct{14b}).
Hybrid inflation in supersymmetric flipped $SU(5)$ predicts the
cosmic microwave anisotropy $\delta T/T$ which is proportional to
$(M/M_P)^2$ (here $M$ denotes the symmetry breaking scale and $M_P
= 2.4\times 10^{18}$ GeV is reduced Planck mass) \ct{15}. Flipped
$SU(5)$ suggests a natural source for the production of Ultra-High
Energy Cosmic Rays (UHECRs) through the decay of superheavy
particles called ``cryptons" which are additional candidates for
Cold Dark Matter (CDM) \ct{16,16a}.

Flipped $SU(5)$ unifies $SU(3)_C$ and $SU(2)_W$ giving the
unification scale $M_{GUT}\sim 10^{16}$ GeV. The discrepancy
between the unification scale $M_{GUT}$ and string scale
$M_{string}\sim 10^{18}$ GeV has only one explanation: there are
extra intermediate symmetries. In Ref.~\ct{17} an $N=1$
supersymmetric flipped $SO(10)$ GUT model was constructed
providing a desired connection with superstrings. This model was
developed in Refs.~\ct{18}. Recently flipped $E_6$ model \ct{19}
was extensively investigated due to the heterotic string theory.
$E_6$ maybe broken down to $SU(3)_C\times SU(2)_L\times
U(1)_X\times U(1)_Z$ (see \ct{17,20}), or $SU(3)_C\times
SU(2)_L\times SU(2)_R\times U(1)$ (see \ct{18,21}).

In the present paper we have given a number of available paths
from the Standard Model (SM) to the $E_6$ unification visually
demonstrating a chain of the flipped models which can follow the
extension of the SM.

The paper is organized as follows. In Sect.~2 we consider the
content of the flipped $E_6$ group of symmetry. In Sect.~3 we
discuss the two possibilities of the extension of the Standard
Model: non-supersymmetric and supersymmetric $SU(3)_C\times
SU(2)_L\times U(1)_X\times U(1)_Z$, originated at the seesaw scale
$M_{SS}$ by heavy right-handed neutrinos. Sect.~4 is devoted to
the general consideration of gauge coupling constants running for
$SU(N)$ and $SO(10)$ gauge groups of symmetry. In Sect.~5 we
present examples of the $E_6, \,SO(10)$ and $SU(5)$ or $SU(4)$
breakdowns towards the Standard Model, depending on the different
contents of the Higgs bosons in the flipped models considered.

\section{Content of the flipped $E_6$}

The first step in the extension of the SM is related with the
left-right symmetry. These models (see \ct{22,22a,22b,22bb}, for
example), where an extra gauge boson $W_R$ mediates charged
right-handed currents, may be embedded in the GUTs $SU(5)$,
$SO(10)$, $E_6$, etc. by different ways.

Among a considerable quantity of articles devoted to the $SU(5)$,
$SO(10)$ and $E_6$ unifications (see, for example, monographs
\ct{22c,22d,22e}, reviews \ct{22f,22g,22i,22k,22l}, articles
\ct{22m,22n,22p,22q,22qq,22rr}, also \ct{7a,7b,7c,7d,7e,7g} and
references therein) we would like to allocate the flipped models
\ct{10,11,12,12a,13,14,14a,14b,15,16,16a,17,18,19,20,21}.

In the present paper we consider the $N=1$ supersymmetric flipped
$E_6$ broken down to $SU(3)_C\times SU(2)_L\times U(1)_Y$ with
intermediate flipped $SO(10)\times U(1)$, $SU(5)\times U(1)$ and
$SU(3)_C\times SU(2)_L\times U(1)_Z\times U(1)_X$ gauge groups of
symmetry. In this model $SU(5)$ (termed 'flipped $SU(5)$') is
distinct from the well-known Georgi--Glashow $SU(5)$ \ct{23}.

The possibility of the different breaking of $E_6$ is a result of
the fact that there are two ways to embed the electric charge
generator in $SU(5) \times U(1)\subset SO(10)$ giving the observed
charges to the quarks and leptons of a family \ct{24}. In the case
of the flipped $SU(5)$, its decomposition into $SU(3)_C\times
SU(2)_L\times U(1)_Z$ is not the same as in the Standard Model,
but leads to the SM only after linearly combining $Z$ and $X$ into
the weak hypercharge $Y$ (see Ref.~\ct{24}): \be
     \frac Y2 = \frac15(X - Z).                       \lb{1}
\ee Here $Z$ is the generator of $SU(5)$ gauge group normalized
for the five-dimensional representation of $SU(5)$ as diag( -1/3,
-1/3, -1/3, 1/2, 1/2). It commutes with $SU(3)_C\times SU(2)_L$.
In the flipped models $SU(3)_C\times SU(2)_L$ is embedded into
$SU(5)$, but $U(1)_Y$ is embedded into $SU(5)\times U(1)_X$.

Three 27-plets of $E_6$ contain three families of quarks and leptons,
including right-handed neutrinos $N_i^c$ (i=1,2,3 is the index of generations).
Matter fields (quarks and leptons)
of the fundamental 27 representation of the flipped $E_6$ decompose under
$SU(5)\times U(1)_X$ subgroup as follows:
\be
        27 \to (10,1) + (\bar 5,2) + (\bar 5, -3) +
          (5,-2) + (1,5) + (1,0).                             \lb{2}
\ee The first and second quantities in the brackets of
Eq.(\ref{2}) correspond to the $SU(5)$ representation and $U(1)_X$
charge, respectively. The SM family which contains the doublets of
left-handed quarks $Q$ and leptons $L$, right-handed up and down
quarks $u^c$, $d^c$, also $e^c$ and right-handed neutrino $N^c$
belong to the $(10,1) + (\bar 5,-3) + (1,5)$ representations of
the flipped $SU(5)\times U(1)_X$. These representations decompose
under \be SU(5)\times U(1)_X \to SU(3)_C\times SU(2)_L\times
U(1)_X\times U(1)_Z \lb{3a} \ee and correspond to ordinary quarks
and leptons with charges: \be
       Q_X = \frac {1}{\sqrt{40}}X \quad {\mbox{and}} \quad
                          Q_Z =\sqrt{\frac 35}Z.             \lb{3}
\ee
We normalize all the flipped $SU(5)$ generators $T_a$ such that the trace of
$T_a^2$ over the 16 fermions in a single quark-lepton generation is given by
$Tr_{16}(T_a^2)=2$, what is consistent with the $U(1)_Y$ charge normalization.
We consider charges $Q_X$ and $Q_Z$ in the units $\frac {1}{\sqrt{40}}$
and $\sqrt{\frac 35}$, respectively, using assignments: $Q_X=X$ and $Q_Z=Z$.
Then for the decomposition (\ref{3a}), we have:
\bea
       (10,1) \to Q = &\left(\begin{array}{c}u\\
                                          d \end{array}\right) &\sim
                         \left(3,2,\frac 16,1\right),\nonumber\\
&d^{\rm\bf c} &\sim \left(\bar3,1,-\frac 23,1\right),\nonumber\\
&N^{\rm\bf c} &\sim \left(1,1,1,1\right).       \lb{4}\\
(\bar 5,-3) \to &u^{\rm\bf c}&\sim \left(\bar 3,1,\frac 13,-3\right),\nonumber\\
L = &\left(\begin{array}{c}e\\
                                             \nu \end{array}\right) &\sim
                         \left(1,2,-\frac 12,-3\right),               \lb{5}\\
(1,5) \to &e^{\rm\bf c} &\sim \left(1,1,1,5\right).\lb{6}
\eea

The remaining representations in Eq.~(\ref{3}) decompose as follows:

\bea
        (5,-2) \to& D&\sim \left(3,1,-\frac 13,-2\right),\nonumber\\
                   h = &\left(\begin{array}{c}h^+\\
                                               h^0 \end{array}\right) &\sim
                         \left(1,2,\frac 12,-2\right).
                                                              \lb{7}\\
    (\bar 5,2) \to &D^{\rm\bf c} &\sim \left(\bar 3,1,\frac 13,2\right),\nonumber\\
                     h^{\rm\bf c} = &\left(\begin{array}{c}h^0\\
                                               h^- \end{array}\right) &\sim
                         \left(1,2,-\frac 12,2\right).              \lb{8}
\eea
The light Higgs doublets are accompanied by coloured Higgs triplets $D,D^{\rm\bf c}$.

The singlet field $S$ is represented by (1,0):
\be
       (1,0) \to S \sim (1,1,2,2).               \lb{9}
\ee
It is necessary to notice that the flipping of our $SU(5)$:
\be
       d^{\rm\bf c} \leftrightarrow u^{\rm\bf c},\quad
N^{\rm\bf c}\leftrightarrow e^{\rm\bf c}, \lb{10}
\ee
distinguishes this group of symmetry from the standard Georgi-Glashow $SU(5)$
\ct{23}.

The multiplets (\ref{4})-(\ref{6}) fit in the 16 spinorial representation
of $SO(10)$:
\be
       F(16) = F(10,1) + F(\bar 5,-3) + F(1,5).     \lb{11}
\ee
Higgs chiral superfields occupy the 10 representation of $SO(10)$:
\be
       h(10) = h(5,-2) + h^c(\bar 5,2).            \lb{12}
\ee
In addition to the three $27_i$ representations of $E_6$, some models may
consider extra $27'$ and ${\ov {27}}'$ representations
with aim to preserve gauge coupling unification. But such models
are out of our investigation.

\section{First steps in the extension of the Standard Model}

We start from the Standard Model (SM).

It is well-known that in the SM the running of all the gauge
coupling constants is well described by the one-loop approximation
of renormalization group equations (RGEs). For energy scale
$\mu\geq M_t$, where $M_t$ is the top quark (pole) mass, we have
the following evolutions for the inverse fine structure constants
$\alpha_i^{-1} = 4\pi/g^2_i$ ($i=1,2,3$ correspond to the $U(1)$,
$SU(2)_L$ and $SU(3)_C$ groups of the SM) which are revised in
Ref.~\ct{25} using updated experimental results \ct{26}:
 \be
      \alpha_1^{-1}(t) = 58.65 \pm 0.02 - \frac{41}{20\pi}t,
                                                           \lb{13}
\ee \be
      \alpha_2^{-1}(t) = 29.95 \pm 0.02 + \frac{19}{12\pi}t,
                                                            \lb{14}
\ee \be    \alpha_3^{-1}(t) = 9.17 \pm 0.20 + \frac{7}{2\pi}t,
                                                             \lb{15}
\ee where $t = \ln(\mu/{M_t})$. In Eq.~(\ref{15}) the value
$\alpha_3^{-1}(M_t)=9.17$ essentially depends on the value
$\alpha_3(M_Z)\equiv \alpha_s(M_Z)=0.118\pm 0.002$ (see \ct{26}),
where $M_Z$ is the mass of $Z$ boson. The inverse constant
$\alpha_3^{-1}(M_t)$ is given by running of $\alpha_3^{-1}(\mu)$
from $M_Z$ up to $M_t$, via the Higgs boson mass $M_H$. We have
used the central value of the top quark (pole) mass: $M_t = 172.6$
GeV given by \ct{26}, and $M_H=130\pm 15$ GeV, in accordance with
the recent experimental data. The accuracy in Eq.~(\ref{15})
corresponds to the accuracy of $M_H$.

Evolutions (\ref{13})\,-(\ref{15}) are shown up to the seesaw
scale $M_{SS}$ in Fig.~1(a), where $x=\log_{10}\mu({\mbox{GeV}})$
and $t=x\ln 10 - \ln M_t$.

However, we have no guarantee for how far the SM will work up the energy
scale. Among different extensions of the SM, we have considered
only two first possibilities:

\begin{itemize}
\item[1.] An extension of the SM to the gauge symmetry group:
 $$SU(3)_C\times SU(2)_L\times U(1)_X\times U(1)_Z,$$
with $U(1)_X$ and $U(1)_Z$ originated at the seesaw scale $M_{SS}$
where heavy (right-handed) neutrinos appear. The consequent
unification of the group  $SU(3)_C\times SU(2)_L\times U(1)_Z$
into the flipped $SU(5)$ at the GUT scale $M_{GUT}$ leads to the
group of symmetry $SU(5)\times U(1)_X$. Such a possibility was
investigated in Ref.~\ct{27}. In this case supersymmetry comes at
the scale $M_{GUT}$ together with the flipped $SU(5)\times
U(1)_X$.

\item[2.] Supersymmetry starts at the scale $M_{SUSY}\sim O$(TeV),
but supersymmetric group $SU(3)_C\times SU(2)_L\times U(1)_X\times
U(1)_Z$, together with the new physics of heavy right-handed
neutrinos, originates at the seesaw scale $M_{SS}$ which is much
larger than $M_{SUSY}$ ($M_{SS} >> M_{SUSY}$). The supersymmetric
group  $SU(3)_C\times SU(2)_L\times U(1)_Z$ gives the unification
into the supersymmetric flipped $SU(5)$ at the GUT scale
$M_{GUT}$, and the supersymmetric group $SU(5)\times U(1)_X$ works
into the region $M_{GUT} \leq \mu \leq M_{SG}$, that is, up to the
next step of unification at the super-GUT scale $M_{SG}$.
\end{itemize}

These two possibilities are presented in Figs.~1,2, where the
first one is given by Fig.~1, while the second one is shown in
Fig.~2.

Here and below we have used the one-loop approximation for RGEs.
The two-loop approximation is not crucial for our idea, because
its contributions are small.

\subsection{Non-supersymmetric extension of the SM}

In this case which was considered in Ref.~\ct{27}, $M_{SUSY}=
M_{GUT}$ and $M_{SS} < M_{GUT}$. The particles content of the
flipped $SU(5)\times U(1)_X$ is given not only by
Eqs.(\ref{4})-(\ref{9}). The five-dimensional (Weinberg-Salam)
Higgs boson breaks $SU(2)_L\times U(1)_Y \to U(1)_{em}$. Also an
adjoint 24-dimensional Higgs boson $A$ ensures the breakdown:
$$SU(5)\times U(1)_X \to SU(3)_C\times SU(2)_L\times U(1)_X\times U(1)_Z.$$
A singlet Higgs field $S$, which can be combined with the Higgs bosons of higher
representations of $SU(5)$, provides the breakdown
$$ SU(3)_C\times SU(2)_L\times U(1)_X\times U(1)_Z \to
    SU(3)_C\times SU(2)_L\times U(1)_Y.$$
In the region $M_t\leq \mu \leq M_{SS}$ we have evolutions of
$\alpha_i^{-1}(t)$ given by Eqs.(\ref{13})-(\ref{15}), but for
$M_{SS}\leq \mu \leq M_{GUT}$ we have a new type of symmetry
$SU(3)_C\times SU(2)_L\times U(1)_X\times U(1)_Z$ with the
following evolutions of the inverse constants $\alpha_i$:
\be
  \alpha_i^{-1}(\mu) = \alpha_i^{-1}(M_{SS}) + \frac{b_i}{2\pi}{\tilde t},
                                           \lb{16}
\ee where $i=3,2,X,Z$ and
\be
              b_X = - \frac {41}{10},\quad  \quad
                      b_Z = - \frac {45}{10},      \lb{17}
\ee and $b_2$, $b_3$ again are given by Eqs.~(\ref{14}) and
(\ref{15}): \be
      b_2 = \frac {19}{6}\quad {\mbox{and}} \quad
                      b_3 = 7.                      \lb{18}
\ee
In Eq.~(\ref{16}):
\be
       \tilde t =\ln(\frac{\mu}{M_{SS}}) = t +
         \ln(\frac{M_t}{M_{SS}})= x\ln 10 - \ln M_{SS}. \lb{18a}
\ee
A new $U(1)_{(B-L)}$ group comes at the seesaw scale $M_{SS}$
and its mixture with $U(1)_Y$ leads to the
following relation (see Refs.~\ct{10,27}):
\be
\alpha_1^{-1}(M_{SS}) = \frac{24}{25}
\alpha_X^{-1}(M_{SS}) + \frac{1}{25}\alpha_Z^{-1}(M_{SS}).
                                                           \lb{19}
\ee
At the scale $M_{GUT}$ we have the unification
$SU(3)_C\times SU(2)_L\times U(1)_Z \to SU(5)$:
\be
  \alpha_Z(M_{GUT}) = \alpha_2(M_{GUT}) = \alpha_3(M_{GUT}) =
  \alpha_{GUT}.       \lb{20}
\ee The GUT scale $M_{GUT}$ is given by the intersection of the
evolutions (\ref{14}) and (\ref{15}) for $\alpha_2^{-1}(t)$ and
$\alpha_3^{-1}(t)$. The evolutions of the inverse constants
$\alpha_i^{-1}(x)$ for $i=1,2,3,X,Z$ in the region  $M_t\leq \mu
\leq M_{GUT}$ are shown in Fig.~1(a).

\subsection{Supersymmetric extension of the SM up to the GUT scale}

In this Subsection we consider the Minimal Supersymmetric Standard Model
(MSSM) when supersymmetry extends the conventional Standard Model from the
scale $M_{SUSY}\sim (O)$ TeV, and gives the following evolutions in the
region $M_{SUSY}\leq \mu \leq M_{SS}$:
\be
  \alpha_i^{-1}(\mu) = \alpha_i^{-1}(M_{SUSY}) + \frac{b_i}{2\pi}{\bar t},
                                           \lb{21}
\ee
with (see Refs.~\ct{28,29}):
\be
         b_1 = - \frac {33}{5}, \quad b_2 = -1, \quad b_3 = 3  \lb{22}
\ee
and $\bar t = \ln(\mu/M_{SUSY})$.

In the region from $M_{SS}$ to $M_{GUT}$ we have a new type of
symmetry --- supersymmetric $SU(3)_C\times SU(2)_L\times
U(1)_X\times U(1)_Z$ with the evolutions ($i=X,Z,2,3$): \be
  \alpha_i^{-1}(\mu) = \alpha_i^{-1}(M_{SS}) + \frac{b_i}{2\pi}{\tilde t},
                                           \lb{23}
\ee
where
\be
      b_X = -\frac{33}{5},\quad b_Z = -9, \quad b_2 = -1, \quad b_3 = 3,
                                            \lb{24}
\ee and $\tilde t$ is given by Eq.~(\ref{18a}).

The mixture of the two $U(1)$ groups -- $U(1)_Y$ and
$U(1)_{(B-L)}$ -- at the seesaw scale $M_{SS}$ leads to the
relation (\ref{19}) for the supersymmetric case.

Flipped $SU(5)$ comes at the GUT scale $M_{GUT}$ given by the new
intersection of the supersymmetric evolutions (\ref{23}) and
(\ref{24}) for $\alpha_2^{-1}(t)$ and $\alpha_3^{-1}(t)$. In
Ref.~\ct{10} the breakdown of $SU(5)\times U(1)_X$ to the MSSM
gauge group is achieved by providing super-large VEVs to the 10
dimensional Higgs $10_H$ and $\ov {10}_{H}$ along the $N^c,\,\bar
N^c$ directions.

The corresponding evolutions of the inverses of the fine structure
constants $\alpha_i^{-1}(x)$ ($i=1,2,3,X,Z$) are presented by
Figs.~2(a,b) for the SM and MSSM in the region $M_t\leq \mu \leq
M_{GUT}$.

\section{Renormalization group equations for gauge couplings
of  the flipped $SU(5)$ and $SO(10)$ }

The renormalization beta-group function $\beta(g)$ has been
calculated by various authors (\ct{30,31,32,33,34,35,36,37}, also
\ct{28} and \ct{29}) for the Yang-Mills theory described by simple
or semi-simple gauge group $G$, including multiplets of fermions
and scalars transforming according to the arbitrary representation
of a group G. Supersymmetric models involve new fields giving new
contributions to the $\beta$-function. In general, considering the
supersymmetric gauge group $G$, we have the following
$\beta$-function in the one-loop approximation:
\be
           \beta_N (g_N) = - b_N \frac{g_N}{16\pi^2}       \lb{25}
\ee
with (see \ct{28})
\be
       - b_N = -3C_2(G) + 2N_g + \sum_i C_2(S_i)d(S_i),    \lb{26}
\ee
where $N_g$ is the number of generations of fermions, scalars $S_i$
belong to representations $R_i$ of the group $G$, $C_2(G)$ is the quadratic
Casimir operator for the adjoint representation, $C_2(S_i)$ is the quadratic
Casimir operators for scalars $S_i$ and $d(S_i)$ is the dimension of the
representation $R_i$.

For $SU(N)$ gauge group $C_2(G)=N,\,\,$ for $SO(N)$ we have
$C_2(G)=\frac 12(N-2)$.

In general, RGE for $SU(N)$ is given by
\be
      - b_N = -3N + 2N_g + \frac 12 N_{vector} + N\cdot N_{adj} +
              \frac 12(N-2)\cdot N_{asym} + \frac 12 (N+2)\cdot N_{sym} + ...
                                                          \lb{27}
\ee
and for $SO(N)$ we have:
\be
   - b_N = -\frac 32(N-2) + 2N_g + \frac 12N_{vector} + \frac 12(N-2)\cdot
N_{adj} + \frac 12(N-2)\cdot N_{asym} + \frac 12(N+2)\cdot N_{sym} + ...,
                                                              \lb{28}
\ee where $N_{vec},\,\,N_{adj},\,\,N_{asym}, \,\,N_{sym}\,$ are
the numbers of (available by models) Higgs bosons belonging to the
vector, adjoint, antisymmetric and symmetric 2nd rank tensor
representations of $SU(N)$ and $SO(N)$ gauge groups, respectively.
We have used the results of Refs.~\ct{33,34}.

Finally, for supersymmetric $SU(5)$ we obtain the following RGE parameter
$b_5$:
\be
     - b_5 = - 15 + 2N_g + \frac 12N_5 + 5N_{24} + \frac 32N_{10}
             + \frac 72N_{15} + ...                      \lb{29}
\ee contained in the evolution of the inverse $SU(5)$ fine
structure constant: \be
  \alpha_5^{-1}(\mu) = \alpha_5^{-1}(M_{GUT}) + \frac{b_5}{2\pi}{t_5},
                                           \lb{30}
\ee
where  $ t_5 = \ln(\mu/M_{GUT})$.
In Eq.~(\ref{29}) $N_5$ is the number of Higgs bosons belonging to the
$5+\bar 5$ representations of $SU(5)$ (minimal value is $N_5=2$),
$N_{24}$ is the number of Higgses belonging to the adjoint representation
of $SU(5)$ (minimal $N_{24}=1$), $N_{10}$ is the number of antisymmetric
(2nd rank tensor) representations $10+\ov {10}$ of $SU(5)$
(minimal $N_{10}=2$),
and $N_{15}$ is the number of symmetric (2nd rank tensor) representations
$15+\ov {15}$ of $SU(5)$ (minimal $N_{15}=2$). We do not consider higher
representations for Higgs bosons: the generalization is trivial and
not necessary for our purposes.

Analogous formula takes place for supersymmetric $SO(10)$ parameter $b_{10}$:
\be
     - b_{10} = - 12 + 2N_g + \frac 12N_{vec} + 4N_{adj} + 4N_{asym}
             + 6N_{sym} + ...                      \lb{31}
\ee contained in the evolution of the inverse $SO(10)$ fine
structure constant: \be
 \alpha_{10}^{-1}(\mu) = \alpha_{10}^{-1}(M_{SG}) + \frac{b_{10}}{2\pi}
{t_{10}},
                                           \lb{32}
\ee
where  $ t_{10} = \ln(\mu/M_{SG})$.
In Eq.(\ref{31}) $N_{vec}$ is the number of Higgs bosons belonging to the
$10+\ov {10}$ representations of $SO(10)$ (minimal value is $N_{vec}=2$),
$N_{adj}$ is the number of Higgses belonging to the adjoint representation 45
of $SO(10)$ (minimal $N_{adj}=1$), $N_{asym}$ is the number of antisymmetric
(2nd rank tensor) representations $45+\ov {45}$ of $SO(10)$ (minimal
$N_{asym}=2$), and $N_{sym}$ is the number of symmetric (2nd rank tensor)
representations $54+\ov {54}$ of $SO(10)$ (minimal $N_{sym}=2$). We
do not consider higher Higgs representations for our investigation.

With aim to break
$$ SO(10) \times U(1)_{X1} \to  SU(5)\times U(1)_{Z1}\times U(1)_{X1}$$
and
$$SU(5)\times U(1)_X \to SU(3)_C\times SU(2)_L\times U(1)_Z\times U(1)_X$$
we have taken the different Higgs boson contents.

\section{Breakdown of the flipped   $SU(5)$  to
the  seesaw scale extensions of the SM and MSSM }

{\bf Case I.} The first possibility to break supersymmetric
$SU(5)\times U(1)_X$ into the non-supersymmetric $SU(3)_C\times
SU(2)_L\times U(1)_X\times U(1)_Z$ was investigated in
Ref.~\ct{27}. In this case $SU(5)\times U(1)_X$ contains three
generations of fermion fields ($N_g=3$, i=1,2,3): \be
      F_i = (10,1), \quad \bar f_i = (\bar 5,-3),\quad l^c_i = (1,5), \lb{33}
\ee
Higgs bosons $h$ and $h^c$ belonging to $5+\bar 5$ representations and Higgs
field $A$ belonging to the adjoint representation of $SU(5)$. In this case, according to
Eq.~(\ref{29}), we have:
\be
          -b_5 = -15 + 6 + 1 + 5 = -3           \lb{34}
\ee
and
\be
                    b_X(SUSY) = -\frac {33}5 = -6.6.      \lb{35}
\ee
Singlet Higgs field $S=(1,0)$ breaks
$$ SU(3)_C\times SU(2)_L\times U(1)_X\times U(1)_Z
\to SU(3)_C\times SU(2)_L\times U(1)_Y.$$
The evolution of $\alpha_5^{-1}(x)$ is shown in Fig.~1(a,b).

At the next seesaw scale $M_{SS1}$ we see the
creation of an additional $U(1)_N$ gauge group which gives a mixture
with $U(1)_X$ and leads to the group of symmetry
$$SU(5)\times U(1)_{Z1}\times U(1)_{X1}.$$ The last one
exists in the region  $M_{SS1}\leq \mu \leq M_{SG}$. Here $M_{SG}$
is the super-GUT scale of the next unification --- $SO(10)$.

Assuming that a new singlet Higgs $S_1$  has the same
$U(1)_{Z1}$ and $U(1)_{X1}$ charges as a singlet $S$,
we have:
\be
   Q_{X1} = 2\frac {1}{\sqrt{40}} =\frac 1{\sqrt{10}} \quad {\mbox{and}} \quad
                          Q_{Z1} = 2\sqrt{\frac 35}.             \lb{36}
\ee This leads to the same slopes for the  running of the inverse
constants
 $\alpha_{X,Z}^{-1}(\mu)$
and $\alpha_{X1,Z1}^{-1}(\mu)$:
\be
    b_X = b_{X1} = -6.6,\qquad b_Z = b_{Z1} = -9.     \lb{37}
\ee
The mixture of the two groups $U(1)_X$ and $U(1)_N$
into $U(1)_{X1}$ and $U(1)_{Z1}$ at the new seesaw scale $M_{SS1}$
gives the following relation analogous to Eq.(\ref{19}):
\be
\alpha_X^{-1}(M_{SS1}) = \frac{24}{25}
\alpha_{X1}^{-1}(M_{SS1}) + \frac{1}{25}\alpha_{Z1}^{-1}(M_{SS1}).
                                                           \lb{38}
\ee As a result, at the super-GUT scale $M_{SG}$ we have the
second unification:
$$SU(5)\times U(1)_{Z1} \to SO(10),$$    and
the group of symmetry $SO(10)\times U(1)_{X1}$ works for $\mu \geq M_{SG}$.

The evolution of the inverse $\alpha_{10}^{-1}(t_{10})$ is given
by \be
     - b_{10} = -12 + 6 + 1 + 4 = - 1,                \lb{39}
\ee
where we have used Eq.~(\ref{31}), inserting $N_{vec}=N_{10}=2$ and
$N_{adj}=N_{45}=1$. Here we assumed the existence of the same
Higgses - vector and adjoint - following the flipped $SU(5)$ content
considered in this Section.

The third seesaw scale $M_{SS2}$ leads to the group of symmetry
$$SO(10)\times U(1)_{X2}\times U(1)_{Z2}$$
with
\be
  b_{X2} = -6.6\quad {\mbox{and}} \quad b_{Z2} = -9, \lb{40}
\ee and with a new condition: \be \alpha_{X1}^{-1}(M_{SS2}) =
\frac{24}{25} \alpha_{X2}^{-1}(M_{SS2}) +
\frac{1}{25}\alpha_{Z2}^{-1}(M_{SS2}).
                                                           \lb{41}
\ee The new unification at the super-super-GUT scale $M_{SSG}$
gives:
$$SO(10)\times U(1)_{Z2} \to E_6$$
with the final symmetry $E_6\times U(1)_{X2}$. The corresponding evolution
is presented in Fig.~1(a,b).

We are unable to predict any seesaw scale, or GUT scales $M_{SG}$
and $M_{SSG}$. In Fig.~1 we have presented an example with
$M_{SG}= 10^{18.3}$ GeV and $M_{SSG}=10^{19}$ GeV.

Fig.~2 presents the {\bf Case I} with supersymmetric seesaw scale
extension of the SM (see Subsection 3.2) and corresponds to the
$M_{SUSY}=10$ TeV.

Here it is pertinent to emphasize that only this type of
unification ({\bf Case I}) conserves the asymptotic freedom: the
running of the inverse gauge coupling constants $\alpha_5^{-1}(t)$
and $\alpha_{10}^{-1}(t)$ shows the asymptotically free behavior.
In general, not every type of unification leads to the asymptotic
freedom at high energies (see below).

{\bf Case II}. In the case II we consider Higgses $5_h + \bar
5_h$, $15_{H'} + \ov {15}_{H'}$ and $A$ belonging to the
24-dimensional adjoint representation of the flipped $SU(5)$, also
$SO(10)$ Higgs bosons $10_h + \ov {10}_h$, $54_{H'} + \ov
{54}_{H'}$ and $A$ belonging to 45-adjoint representation. Then we
have: \be
          -b_5 = -15 + 6 + 1 + 5 + 7 = 4           \lb{47}
\ee
and
\be
     -b_{10} = -12 + 6 + 1 + 4 + 12 = 11,                \lb{48}
\ee
with $N_{vec}=2$, $N_{sym}=2$ and $N_{adj}=1$ in Eqs.~(\ref{29}) and
(\ref{31}).

The asymptotically unfree evolutions of $\alpha_5^{-1}(x)$ and
$\alpha_{10}^{-1}(x)$ are presented in Fig.~3 and Fig.~4. All
these figures are given for $M_{SS}=10^{11}$ GeV. Fig.~3 shows the
final unification $E_6\times U(1)_{X2}$ with
$\alpha_6^{-1}(M_{SSG})\approx 42$ for $M_{SUSY}=M_{GUT}$
(non-supersymmetric seesaw scale extension of the SM), and
$\alpha_6^{-1}(M_{SSG})\approx 21$ for $M_{SUSY}=10$ TeV.

However, special case of parameters given by Fig.~4 may lead to
the pure $E_6$ unification. Fig.~4 demonstrates such an example
with $\alpha_6^{-1}(M_{SSG})\approx 18$ for $M_{SUSY}=1$ TeV.

{\bf Case III}. This case considers Higgs bosons $5_h + \bar 5_h$,
$10_H + \ov {10}_H$, $15_{H'} + \ov {15}_{H'}$ and adjoint $A$ of
the flipped $SU(5)$, and correspondingly Higgs bosons $10_h + \ov
{10}_h$, $45_H + \ov {45}_H$, $54_{H'} + \ov {54}_{H'}$ and
adjoint $A$ of the flipped $SO(10)$. Such a consideration gives:
\be
          -b_5 = -15 + 6 + 1 + 5 + 3 + 7 = 7           \lb{49}
\ee
and
\be
     -b_{10} = -12 + 6 + 1 + 4 + 8 + 12 = 19,                \lb{50}
\ee
with $N_{vec}=2$, $N_{asym}=2$, $N_{sym}=2$ and $N_{adj}=1$ inserted
in Eqs.~(\ref{29}) and (\ref{31}).

The evolutions of $\alpha_5^{-1}(x)$ and $\alpha_{10}^{-1}(x)$ are
presented in Fig.~5(a,b ). They are asymptotically unfree.

The supersymmetric seesaw scale extension of the SM always gives
only $E_6$ final unification at the scale $M_{SSG}$. Fig.~5(a,b)
presents this case giving $\alpha_6^{-1}(M_{SSG})\approx 21$ for
$M_{SUSY}=10$ TeV.

For all cases we have: \be
    b_X = b_{X1}  = b_{X2} = -6.6 \quad {\mbox{and}} \quad
    b_Z = b_{Z1} = b_{Z2} = -9.     \lb{51}
\ee
Also the conditions (\ref{38}) and (\ref{41}) take place.

Adding higher representations of Higgs bosons to the
representations $5_h + \bar 5_h$, $10_H + \ov {10}_H$ and adjoint
$A$ of $SU(5)$ we always have asymptotically unfree behavior at
high energies $\mu \geq M_{GUT}$. But in these cases we must take
into account the next loops approximation.

{\bf Case IV}. We can assume that the following supersymmetric
left-right symmetry \ct{22,22a,22b,22bb} originates at the seesaw
scale $M_{SS}$: \be
 SU(3)_C\times SU(2)_L\times
SU(2)_R\times U(1)_X\times U(1)_Z. \lb{7x} \ee Here we see
additional groups $SU(2)_R$ and $U(1)_{(B-L)}$ originated at the
scale $M_{SS}$. The group $U(1)_{(B-L)}$ is mixed with the gauge
group $U(1)_Y$ leading to the product $U(1)_X\times U(1)_Z$.

Considering the running for the supersymmetric group (\ref{7x}) in
the region $\mu\ge M_{SS}$ we have the renormalization scale
$M_{ren} = M_{SS}$ and the following slopes: \be
       b_X = b_1 = - 6.6, \quad b_Z = - 9, \quad b_3 = 3.
                                          \lb{9x} \ee
Also  the running of $SU(2)_L\times SU(2)_R$ in the same region of
$\mu$ is given by the slope: \be
                 b_{22} = - 2,   \lb{10x} \ee
and we have the following evolution: \be
 \alpha_{22}^{-1}(\mu) = \alpha_{22}^{-1}(M_R) + \frac{1}{\pi}\ln\frac{\mu}{M_R},
                                          \lb{10ax} \ee
where \be
         \alpha_{22}^{-1}(M_R) = \alpha_2^{-1}(M_R).
                                                       \lb{10bx} \ee
The next step is an assumption that the group $SU(4)_C\times
SU(2)_L\times SU(2)_R$ by Pati and Salam \ct{22} originates at the
scale $M_4$ giving the following extension of the group
(\ref{7x}): \be
 SU(3)_C\times SU(2)_L\times
SU(2)_R\times U(1)_X\times U(1)_Z \to  SU(4)_C\times SU(2)_L\times
SU(2)_R\times U(1)_Z.  \lb{11x} \ee The scale $M_4$ is given by
the intersection of $SU(3)_C$ with $U(1)_X$: \be
        \alpha_3^{-1}(M_4) = \alpha_X^{-1}(M_4).   \lb{12x} \ee
In the MSSM we have Eq.~(\ref{27}) for the slope $b_N$ of the
$SU(N)$ group. Considering only the minimal content of scalar
Higgs fields, e.g. quartets $4+\bar 4$, we have $N_{vector} = 2$
and obtain the following slope for the running of
$\alpha_4^{-1}(\mu)$: \be
       b_4 = 3\cdot 4 - 6 - 1 = 5.  \lb{14x}     \ee
Now the evolution with $M_{ren} = M_4$ gives: \be
 \alpha_4^{-1}(\mu) = \alpha_4^{-1}(M_4) +
 \frac{5}{2\pi}\ln\frac{\mu}{M_4}.
                                          \lb{14ax} \ee
This is the running for the symmetry group $SU(4)$.

The intersection of $\alpha_4^{-1}(\mu)$ with the running of
$\alpha_{22}^{-1}(\mu)$ leads to the scale $M_{GUT}$ of the
$SO(10)$-unification: \be
        SU(4)_C\times SU(2)_L \times SU(2)_R \to SO(10),
                                             \lb{15x}     \ee
and we obtain the value of $M_{GUT}$ from the relation: \be
 \alpha_4^{-1}(M_{GUT}) = \alpha_{22}^{-1}(M_{GUT}).   \lb{16x} \ee
Then we see the evolution of the $SO(10)$ inverse gauge constant
$\alpha_{10}^{-1}(\mu)$, which runs from the scale $M_{GUT}$ up to
the scale $M_{SGUT}$ of the super-unification $E_6$: \be
     SO(10)\times U(1)_Z \to E_6.    \lb{17x}     \ee
The slope of this running is $b_{10}$.

In general, for the $SO(N)$ group we have the slope (\ref{28}).

Calculating the $SO(10)$-slope we must consider not only vectorial
Higgs fields $N_{vector} = 2$, but also $N_{adjoint} = 1$, because
the appearance of right-handed particles is impossible without
adjoint Higgs field (see explanation in Ref.~\ct{27}). As a
result, we obtain from Eq.~(\ref{28}) the following
$SO(10)$-slope: \be
         b_{10} = 12 - 6 - 1 - 4 = 1,
                                 \lb{19x} \ee
 which gives the following running of $\alpha_{10}^{-1}(\mu)$: \be
 \alpha_{10}^{-1}(\mu)
 = \alpha_{10}^{-1}(M_{GUT}) +
 \frac{1}{2\pi}\ln {\frac{\mu}{M_{GUT}}},
                                              \lb{20x} \ee
valid up to the super-GUT scale $M_{SGUT}$ of the
$E_6$-unification.

Finally, as a result of this investigation, one can envision the
following symmetry breaking chain:
$$
     E_6 \to SO(10)\times U(1)_Z\to SU(4)_C\times SU(2)_L \times SU(2)_R\times U(1)_Z\to $$
     $$SU(3)_C\times SU(2)_L \times SU(2)_R\times U(1)_X\times U(1)_Z
     \to  SU(3)_C\times SU(2)_L\times U(1)_Y. $$
All evolutions of the corresponding inverse fine structure
constants are given in Figs.~6(a,b).

\newpage

\section{Conclusions}

In the present paper we have considered a number of possibilities
how Nature can choose its path from the Standard Model to the
Planck scale. We have investigated a chain of the flipped models:
$$
 SU(3)_C\times SU(2)_L\times U(1)_Y \to
 SU(3)_C\times SU(2)_L\times U(1)_X \times U(1)_Z
\to
$$ $$SU(5)\times U(1)_X \to  SU(5)\times U(1)_{Z1}
\times U(1)_{X1} \to SO(10) \times U(1)_{X1} \to $$ $$ SO(10) \times
U(1)_{Z2}\times U(1)_{X2} \to  E_6\times U(1)_{X2}\quad {\mbox{or}}\quad E_6,
$$
which contains three seesaw scales $M_{SS}$, $M_{SS1}$, $M_{SS2}$
and is ended by the flipped  $E_6\times U(1)$ or $E_6$ final
unification at the scale $M_{SSG}\sim 10^{18}$ GeV.

In this investigation we have showed that different extensions of
the SM, supersymmetric and non-supersymmetric, may lead to the
final unification $E_6\times U(1)$ or $E_6$ at the Planck scale.
These different paths essentially depend on the fact whether the
flipped unifications $SU(5)$ and $SO(10)$ show asymptotically free
or asymptotically unfree behavior of gauge couplings. Such a
behavior is connected with a number of representations of Higgs
bosons, providing the breaking of the flipped  $SO(10)$ and
$SU(5)$ down to the SM.

Also we have considered the possibility of the existence of the
chain with a left-right symmetry:
$$
SU(3)_C\times SU(2)_L\times U(1)_Y \to SU(3)_C\times SU(2)_L
\times SU(2)_R\times U(1)_X\times U(1)_Z$$ $$ \to SU(4)_C\times
SU(2)_L \times SU(2)_R\times U(1)_Z$$ $$ \to SO(10)\times U(1)_Z
\to E_6.
$$
We are not able to predict all seesaw scales and GUT scales. Only
$M_{GUT}$ is fixed by the intersection of evolutions
$\alpha_2^{-1}(x)$ and $\alpha_3^{-1}(x)$ for all cases. But we
are sure that future investigations in Cosmology and Astrophysics
can fix Grand Unification Theory with its parameters.

This paper is devoted to the investigation of different paths from
the SM to the Planck scale physics. Of course, it is far from all
possibilities.  For example, we have omitted:

\begin{itemize}

\item[1.] The extension of the SM with the Two Higgs Doublet Model
(2HDM) \ct{38,39,40}.

\item[2.] The Split Supersymmetry \ct{41}.

\item[3.] The extension of the SM with Family replicated gauge
group models. This very interesting possibility to consider
anomaly free family replicated theories was first suggested in
Refs.~\ct{42,42a}, where the model with gauge group of symmetry
$[SU(3)\times SU(2)\times U(1)]^3$ was developed as an extension
of the SM. A lot of successful results were obtained in this
approach (see reviews \ct{4a,44,45}).

\item[4.] The next step to consider family replicated model
$[SU(5)]^3$ was suggested in Refs.~\ct{46,47,48,49}, and the model
$[SO(10)]^3$ was investigated in Ref.~\ct{50} (see also
Refs.~\ct{48}). Ref.~\ct{51} is devoted to the $[E_6]^3$
unification. In general, the possibility to investigate the
flipped family replicated models is very attractive. We have left
this type of models for future investigations.
\end{itemize}

Recently the $E_8$-unification of the graviweak (see \ct{7x,7y})
and strong interactions was suggested in Ref.~\ct{7z} with the
breakdown $E_8\to E_6\times SU(3)$. Theory predicts the existence
of three generations of the $E_6$-subgroups.

\section{Acknowledgements}

L.V.L. deeply thanks the Institute of Mathematical Sciences
(Chennai, India) and personally Prof.~N.D.~Hari Dass for wonderful
hospitality and financial support.

We thank D.L.~Bennett and H.B.~Nielsen for useful discussions.

\newpage

\newpage

\begin{figure}
\begin{center}
\includegraphics[height=180mm,keepaspectratio=true]{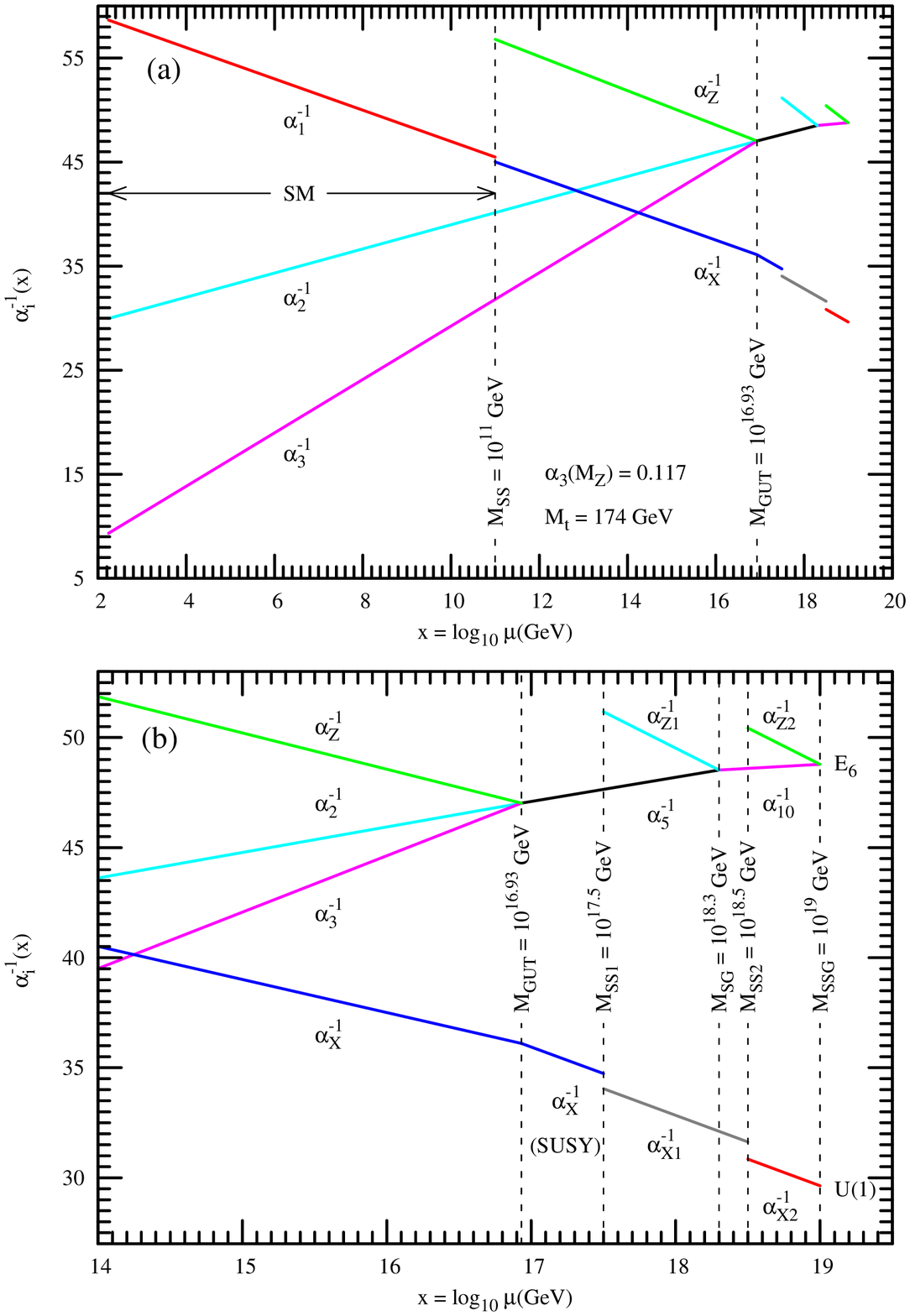}
\caption{The running of the inverse gauge coupling constants
$\alpha_i^{-1}(x)$ for $i=1,2,3,X,Z,X1,Z1,X2,Z2,5,10$ in the {\bf
Case I} which describes the breakdown of the flipped $SU(5)$ to
the non-supersymmetric $SU(3)_C\times SU(2)_L\times U(1)_X\times
U(1)_Z$ gauge group of symmetry with Higgs bosons $5_h + \bar 5_h$
and 24-dimensional adjoint A. This case shows an asymtotically
free behavior of $\alpha_5^{-1}(x)$ and $\alpha_{10}^{-1}(x)$ and
gives the final unification  $E_6\times U(1)$ with
$\alpha_6^{-1}(M_{SSG})\approx 49$. Fig.~1(a) presents the running
from $M_t$ up to the super-super-GUT scale $M_{SSG}\sim 10^{19}$
GeV. Fig.~1(b) gives the same evolutions from $\mu=10^{14}$ GeV. }
\end{center}
\label{fig1}
\end{figure}

\newpage

\begin{figure}
\begin{center}
\includegraphics[height=180mm,keepaspectratio=true]{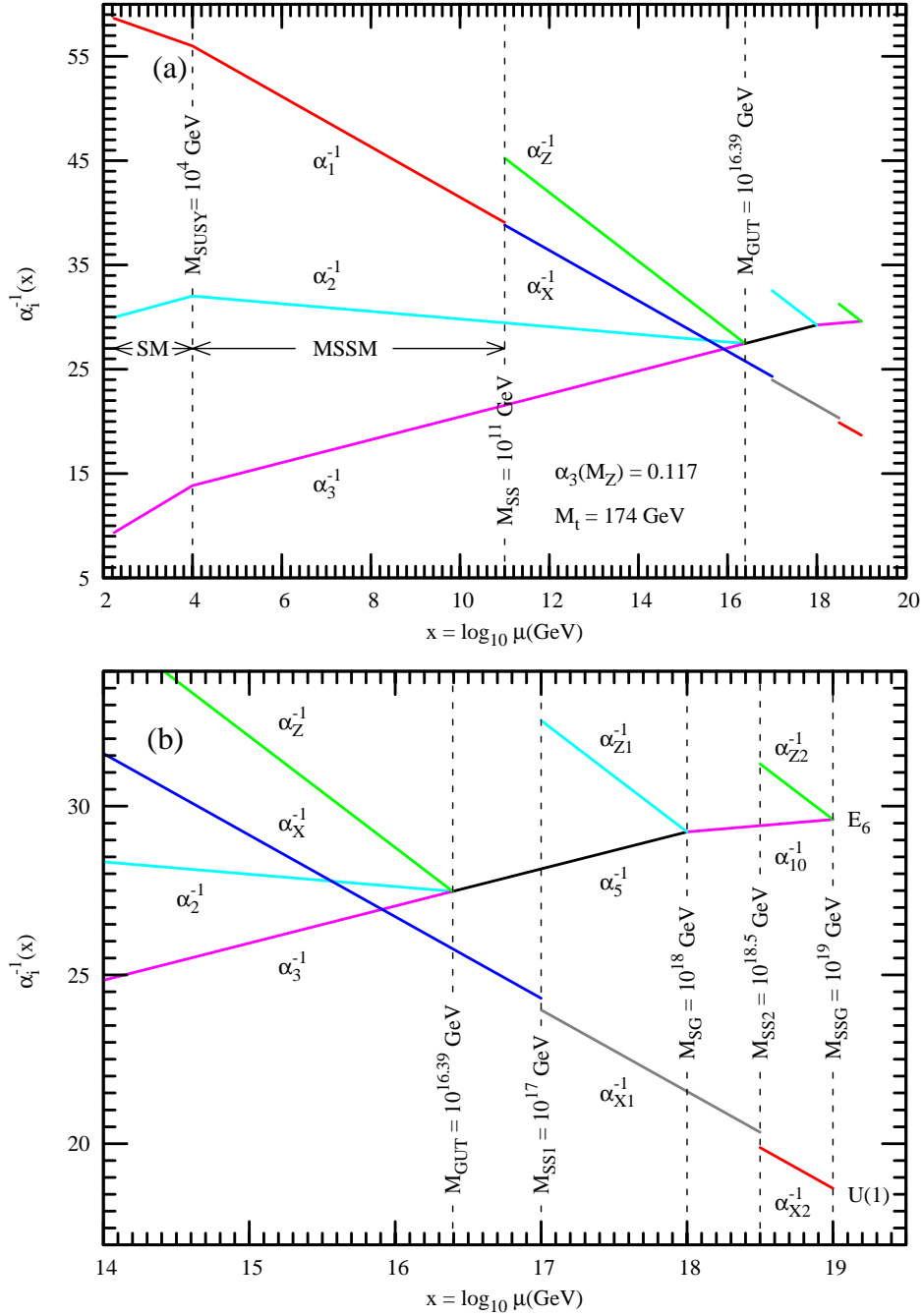}
\caption{The running of the inverse gauge coupling constants
$\alpha_i^{-1}(x)$ for $i=1,2,3,X,Z,X1,Z1,X2,Z2,5,10$ in the {\bf
Case I} corresponding to the breakdown of the flipped $SU(5)$ to
the supersymmetric (MSSM) $SU(3)_C\times SU(2)_L\times
U(1)_Z\times U(1)_X$ gauge group of symmetry with Higgs bosons
belonging to $5_h + \bar 5_h$ and adjoint representations of
$SU(5)$. This example shows an asymtotically free behavior of
$\alpha_5^{-1}(x)$ and $\alpha_{10}^{-1}(x)$. The final
unification is $E_6\times U(1)$ with
$\alpha_6^{-1}(M_{SSG})\approx 30$ for $M_{SUSY}=10$ TeV and
$M_{SS}=10^{11}$ GeV. Fig.~2(a) presents the region of energies
$M_t\leq \mu \leq M_{SSG}$. Fig.~2(b) gives the region of energies
$\mu\geq 10^{14}$ GeV.}
\end{center}
\label{fig2}
\end{figure}

\newpage

\newpage

\begin{figure}
\begin{center}
\includegraphics[height=150mm,keepaspectratio=true,angle=-90]{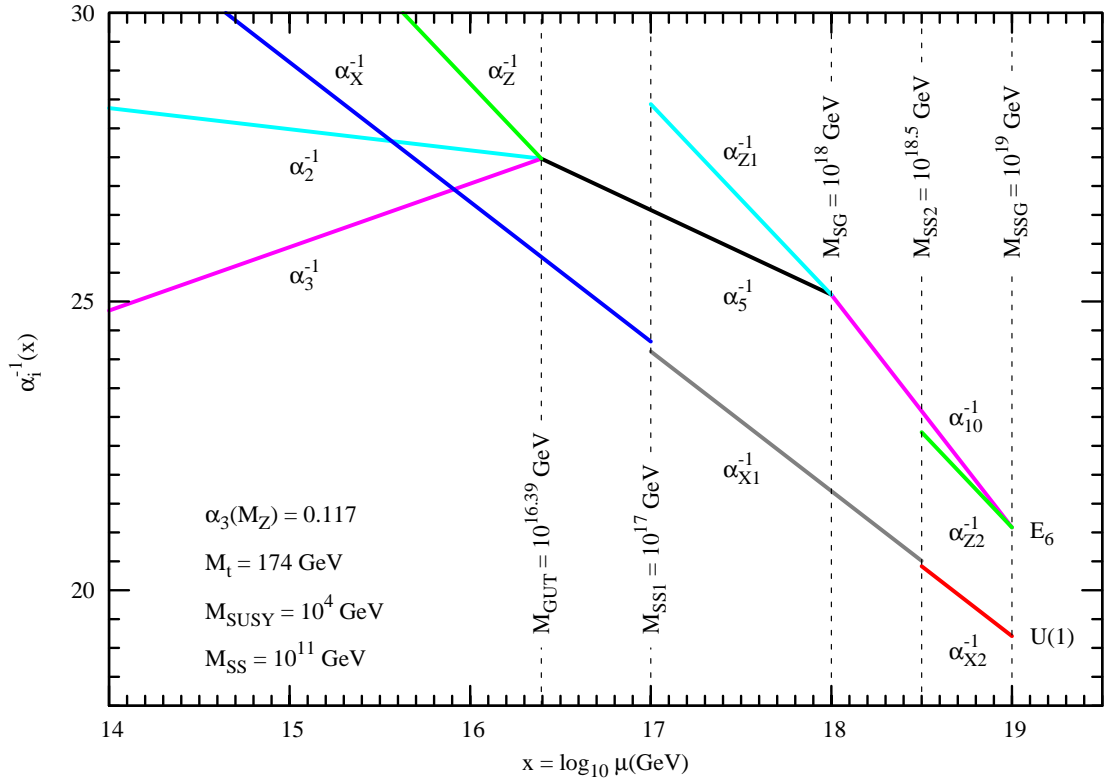}
\caption{The running of the inverse gauge coupling constants
$\alpha_i^{-1}(x)$ with $i=2,3,X,Z,X1,Z1,X2,Z2,5,10$ for the {\bf
Case II} in the region of energies $\mu\geq 10^{14}$ GeV. This
case corresponds to the breakdown of the flipped $SU(5)$ to the
supersymmetric (MSSM) $SU(3)_C\times SU(2)_L\times U(1)_Z\times
U(1)_X$ gauge group of symmetry with Higgs bosons belonging to
$5_h + \bar 5_h$, $15_{H'} + \ov {15}_{H'}$ and adjoint A
representations of $SU(5)$. This example shows an asymtotically
unfree behavior of $\alpha_5^{-1}(x)$ and $\alpha_{10}^{-1}(x)$.
The final unification is $E_6\times U(1)$ with
$\alpha_6^{-1}(M_{SSG})\approx 21$ for $M_{SUSY}=10$ TeV and
$M_{SS}=10^{11}$ GeV.}
\end{center}
\label{fig3}
\end{figure}

\newpage

\begin{figure}
\begin{center}
\includegraphics[height=150mm,keepaspectratio=true,angle=-90]{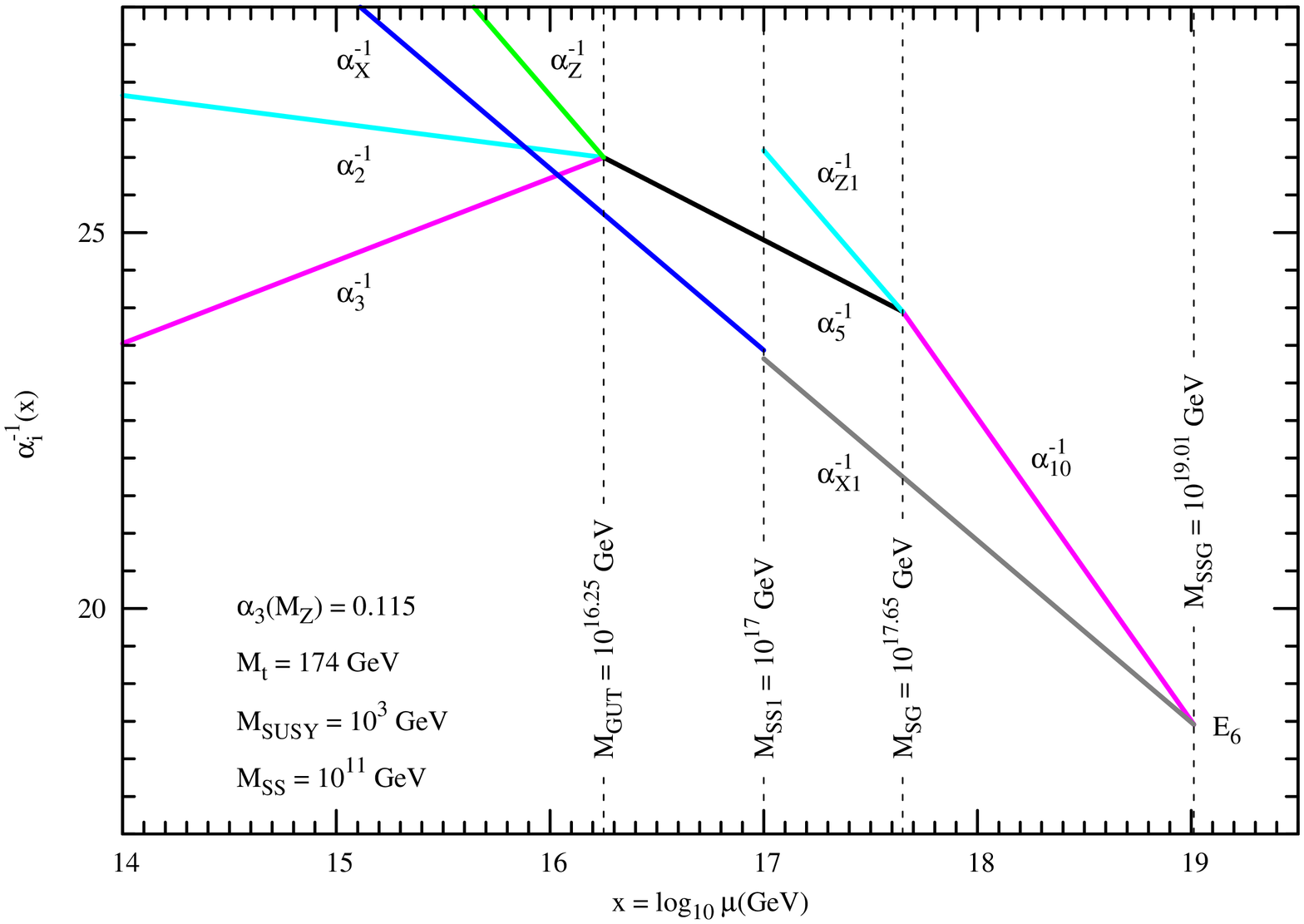}
\caption{The running of the inverse gauge coupling constants
$\alpha_i^{-1}(x)$ with $i=2,3,X,Z,X1,Z1,5,10$ for the {\bf
special Case II} in the region of energies $\mu\geq 10^{14}$ GeV.
The Case II corresponds to the breakdown of the flipped $SU(5)$ to
the supersymmetric (MSSM) $SU(3)_C\times SU(2)_L\times
U(1)_Z\times U(1)_X$ gauge group of symmetry with Higgs bosons
belonging to $5_h + \bar 5_h$, $15_{H'} + \ov {15}_{H'}$ and
adjoint A representations of $SU(5)$. The figure gives an
asymtotically unfree behavior of $\alpha_5^{-1}(x)$ and
$\alpha_{10}^{-1}(x)$. This example shows that for a special set
of parameters the final unification is $E_6$ with
$\alpha_6^{-1}(M_{SSG})\approx 18.5$ for $M_{SUSY}=1$ TeV and
$M_{SS}=10^{11}$ GeV.}
\end{center}
\label{fig4}
\end{figure}

\newpage

\begin{figure}
\begin{center}
\includegraphics[height=180mm,keepaspectratio=true]{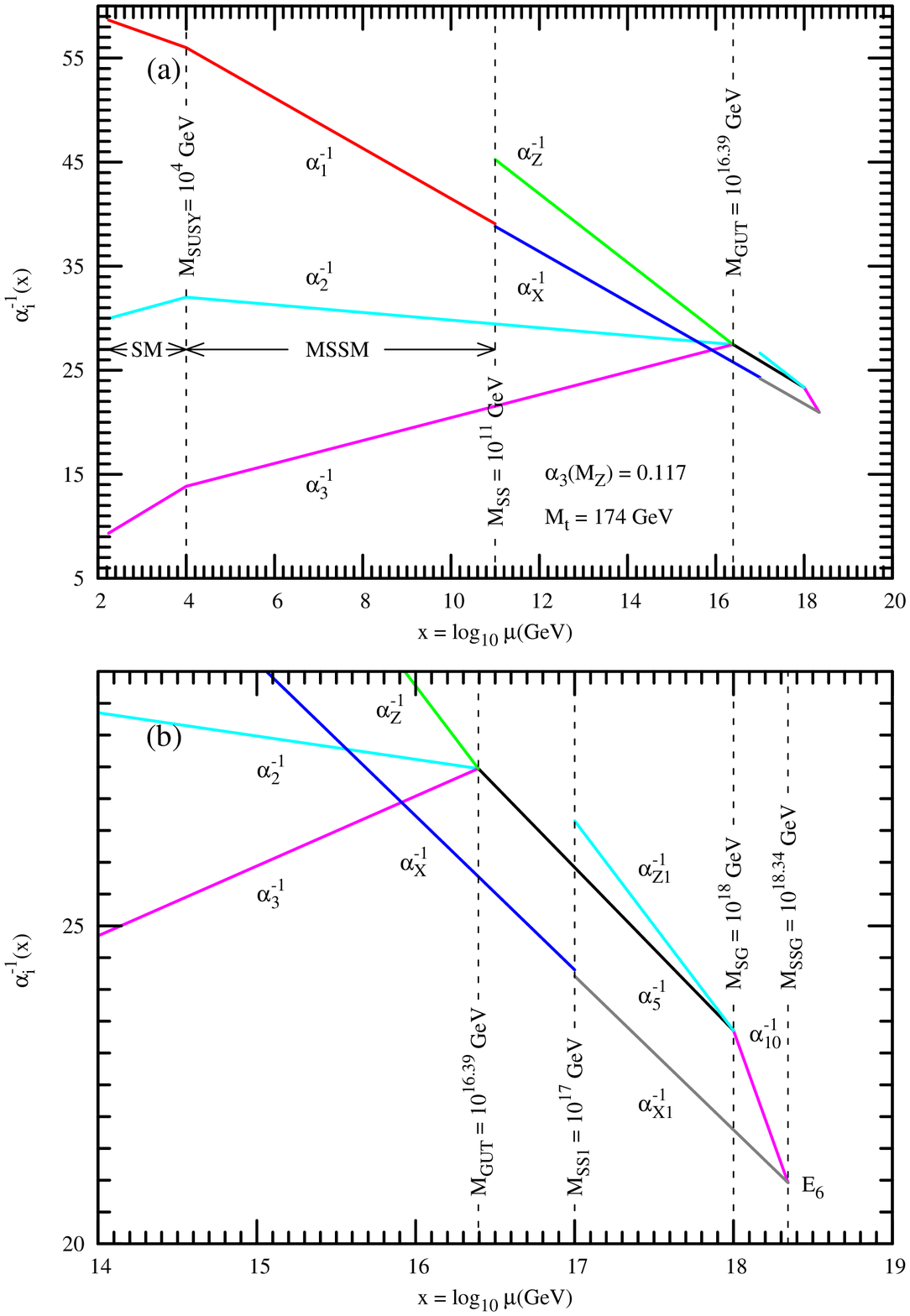}
\caption{The running of the inversed gauge coupling constants
$\alpha_i^{-1}(x)$ for $i=1,2,3,X,Z,X1,Z1,5,10$ in the {\bf Case
III} corresponding to the breakdown of the flipped $SU(5)$ to the
supersymmetric (MSSM) $SU(3)_C\times SU(2)_L\times U(1)_Z\times
U(1)_X$ gauge group of symmetry with Higgs bosons belonging to
$5_h + \bar 5_h$, $10_H + \ov {10}_H$, $15_{H'} + \ov {15}_{H'}$
and adjoint A representations of $SU(5)$. This example shows an
asymtotically unfree behaviour of $\alpha_5^{-1}(x)$ and
$\alpha_{10}^{-1}(x)$. The final unification is $E_6$ with
$\alpha_6^{-1}(M_{SSG})\approx 21$ for $M_{SUSY}=10$ TeV and
$M_{SS}=10^{11}$ GeV. Fig.~4(a) describes the region of energies
$M_t\leq \mu \leq M_{SSG}$. Fig.~4(b) is given for the region of
energies $\mu\geq 10^{14}$ GeV.}
\end{center}
\label{fig5}
\end{figure}

\clearpage\newpage \bfi \centering
\includegraphics[height=100mm,keepaspectratio=true,angle=0]{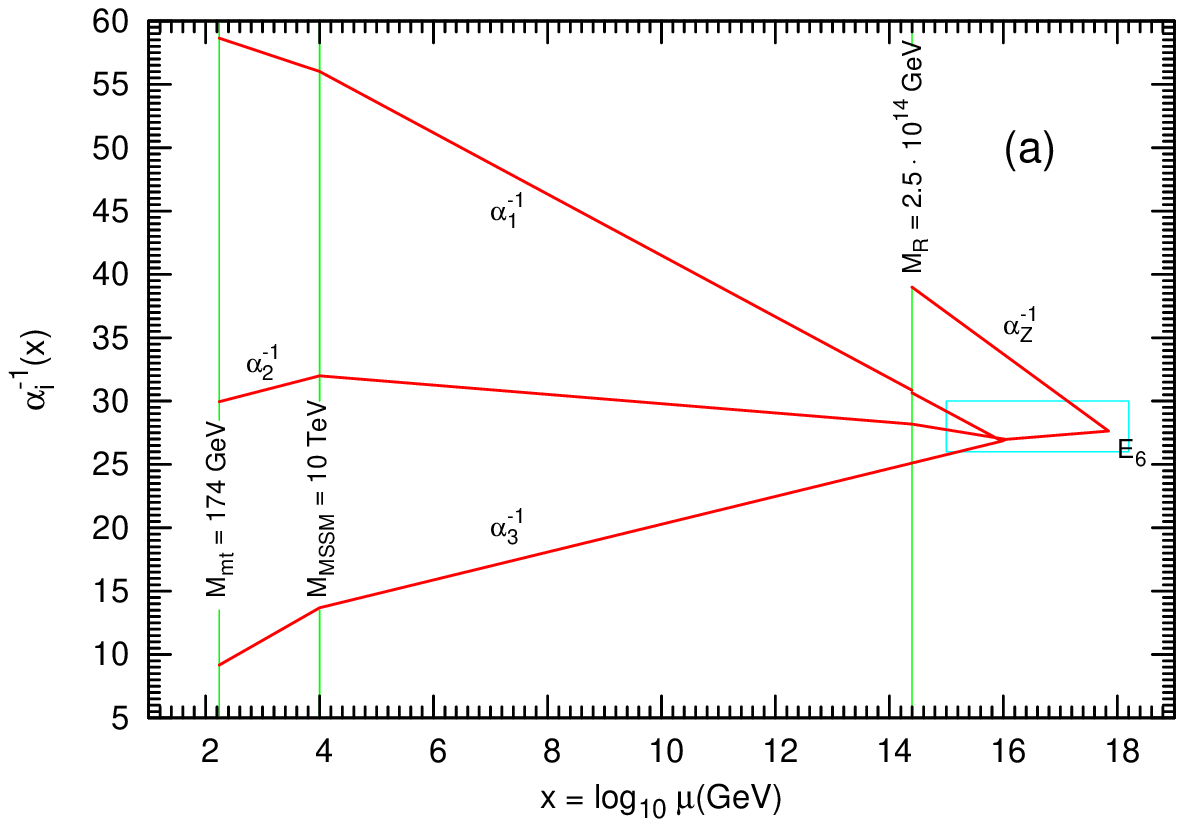}
\includegraphics[height=100mm,keepaspectratio=true,angle=0]{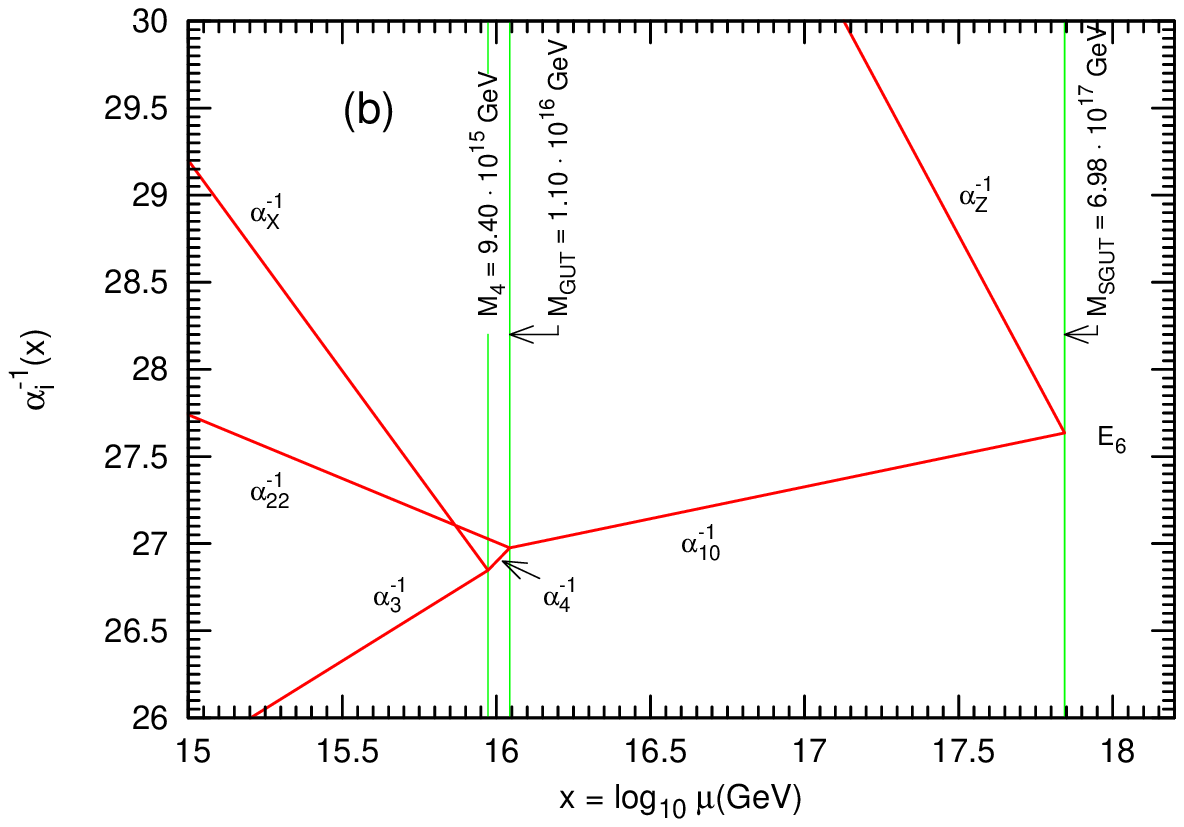}
\caption {Figure (a) presents the running of the inverse coupling
constants $\alpha_i^{-1}(x)$ from the SM up to the $E_6$
unification for SUSY breaking scale $M_{SUSY}= 10$ TeV and seesaw
scale $M_R=2.5\cdot 10^{14}$ GeV. This {\bf Case IV} including a
left-right symmetry gives: $M_{E6}= 6.98\cdot 10^{17}$ GeV and
$\alpha_{E6}^{-1}\approx 27.64$. (b) is same as (a), but zoomed in
the scale region $10^{15}$ GeV up to the $E_6$ unification to show
the details.} \label{fig6} \efi

\end{document}